\documentclass[%
 reprint,
 amsmath,amssymb,
 aps,
]{revtex4-2}

\usepackage{graphicx}% Include figure files
\usepackage{dcolumn}% Align table columns on decimal point
\usepackage{bm}% bold math
\usepackage[colorlinks=true, allcolors=blue]{hyperref}
\begin{document}

\preprint{APS/123-QED}

\title{Comment on “Spin-trap isomers in deformed, odd-odd nuclei in the light rare-earth region near N = 98”}% Force line breaks with \\

\author{N. Susshma, S. Deepa, K. Vijay Sai \& R. Gowrishankar}
 \email{rgowrishankar@sssihl.edu.in}
\affiliation{ Department of Physics, Sri Sathya Sai Institute of Higher Learning, Prashanthi Nilayam, India}%

\date{\today}% It is always \today, today,

\begin{abstract}
Mass spectrometry studies of  odd-odd light rare-earth nuclei by Orford \textit{et al.} [\href{https://link.aps.org/doi/10.1103/PhysRevC.102.011303}{Phys. Rev. C 102, 011303(R) (2020)}] suggested the existence of new isomers in the neutron rich isotopes $^{162}$Tb and $^{164}$Tb. More recently, Stryjczyk \textit{et al.} [\href{https://link.aps.org/doi/10.1103/PhysRevC.111.049801}{Phys. Rev. C 111, 049801 (2025)}] commented on the former, citing inconsistencies between the available experimental data and the proposed presence of an isomer in $^{162}$Tb. 
To further examine the possibility of isomeric states in $^{162,164}$Tb, we employed the well-tested empirical Two Quasiparticle Rotor Model to construct their low-lying level structure.  
The resulting level schemes support the potential existence of low-lying isomeric states in both isotopes and we propose their corresponding spin-parities, orbital configurations, and excitation energies.

\end{abstract}

\maketitle

%\tableofcontents

Orford \textit{et al.} \cite{Orford2020} predicted the existence of low-lying isomers in $^{162}$Tb (J$^\pi$=4$^-$) and $^{164}$Tb (J$^\pi$=2$^+$) from their mass spectroscopy studies. They interpreted these isomers as `spin-traps' and proposed their configurations and energies. 
In a recently published comment on Orford \textit{et al.} \cite{Orford2020}, Stryjczyk \textit{et al.} \cite{Stry2025} proposed that in $^{162}$Tb the mass observed for the J$^\pi$=4$^-$ state is for the ground state J$^\pi$=1$^-$ and the other observed mass could be a molecular contaminant, citing the trend of the S$_{2n}$ curves and the $\beta$-decay reports \cite{Schima66,KAWADE77}. The current evaluated data sheets for $^{162}$Tb \cite{ENSDF}, lists the J$^\pi$= 4$^-$ level with two different energies: E$_x$=216 keV adopted from Burke \textit{et al.} \cite{BURKE2007} and E$_x$=284 keV from Orford \textit{et al.} \cite{Orford2020}. On the other hand, there are no experimental data \cite{ENSDF} available on $^{164}$Tb except its ground state.  
We have carried out an independent study to construct the low-lying level structures of $^{162,164}$Tb investigating the possible existence of isomers. To this end, we have employed the well-tested semi-empirical Two Quasiparticle Rotor Model (TQRM) \cite{Jain90,Jain98} to deduce the physically admissible 2qp bandheads with E$_x<$ 500 keV.

\begin{figure}[!t]
\centering
\includegraphics[width= 17.0 cm,angle=0,scale=0.50,trim=0.5cm 2.2cm 3.0cm 0.5cm,clip=true]{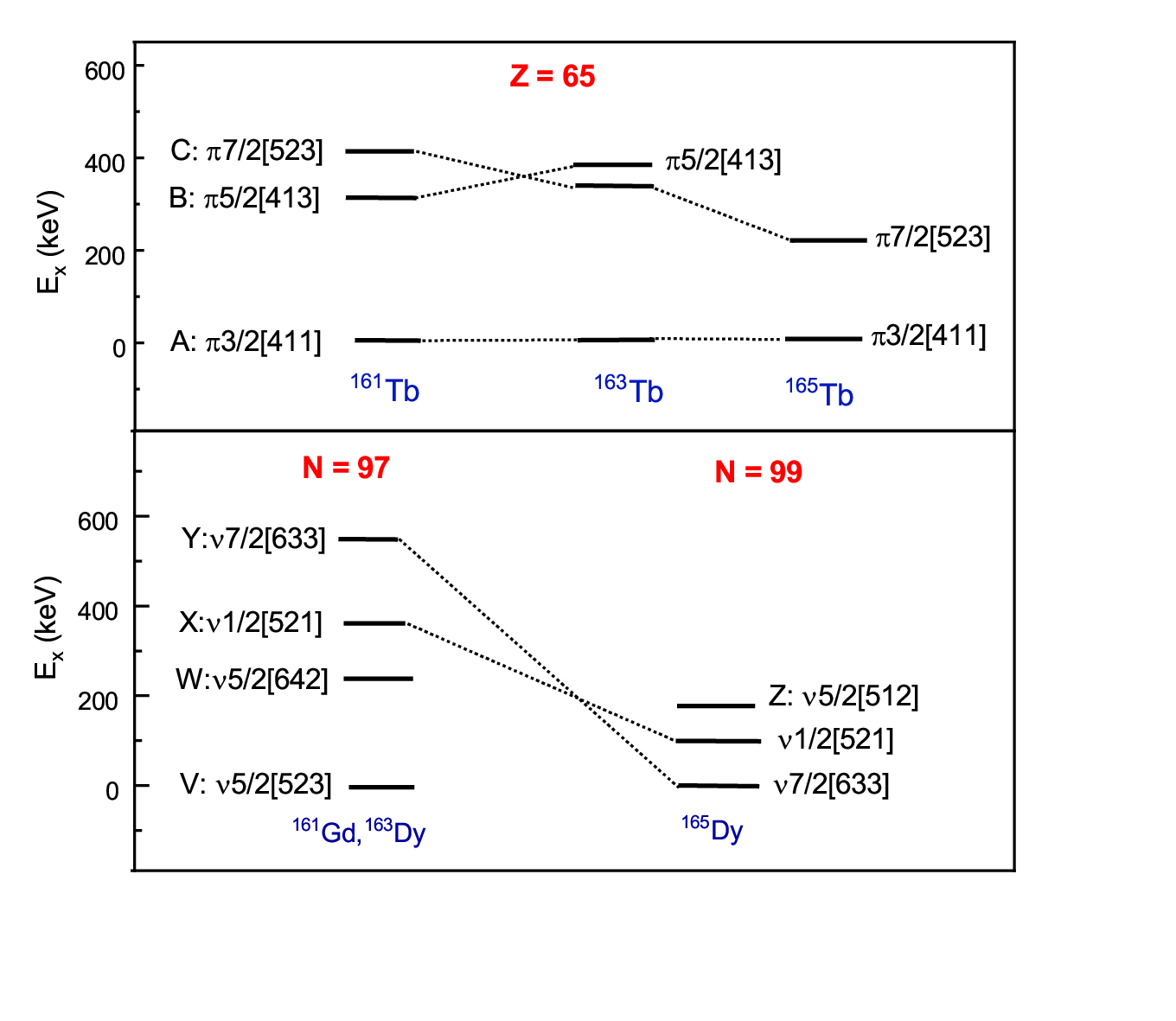}
\caption{Single particle orbital energy systematics of proton orbitals in Z=65 isotopes and neutron orbitals in N=97 and N=99 isotones in the mass regions A=160-165.} \label{fig1}
\end{figure}

Semi-empirical TQRM calculations were carried out following the three standard steps. First, the systematics of the lowest-lying single particle (1qp) proton and neutron orbital energies in the neighboring isotopes and isotones of $^{162}$Tb and $^{164}$Tb were mapped. The systematics of the 1qp orbitals for $^{162}$Tb is plotted in Fig. \ref{fig1}. Following this, spin-parities of the 2qp bandheads (Gallagher-Moszkowski (GM) doublets) formed by the coupling of these 1qp orbitals were determined. The energies of the 2qp states thus formed were calculated from the following TQRM expression \cite{Jain90,Jain98}.

\begin{equation}
     E(\Omega_p,\Omega_n)= E_0+E_p(\Omega_p)+E_n(\Omega_n)+E_{rot} +< V_{pn} > 
\end{equation}

\begin{figure}[!t]
\centering
\includegraphics[width= 17.0 cm,angle=0,scale=0.50,trim=1.2cm 0.5cm 2.7cm 0.6cm,clip=true]{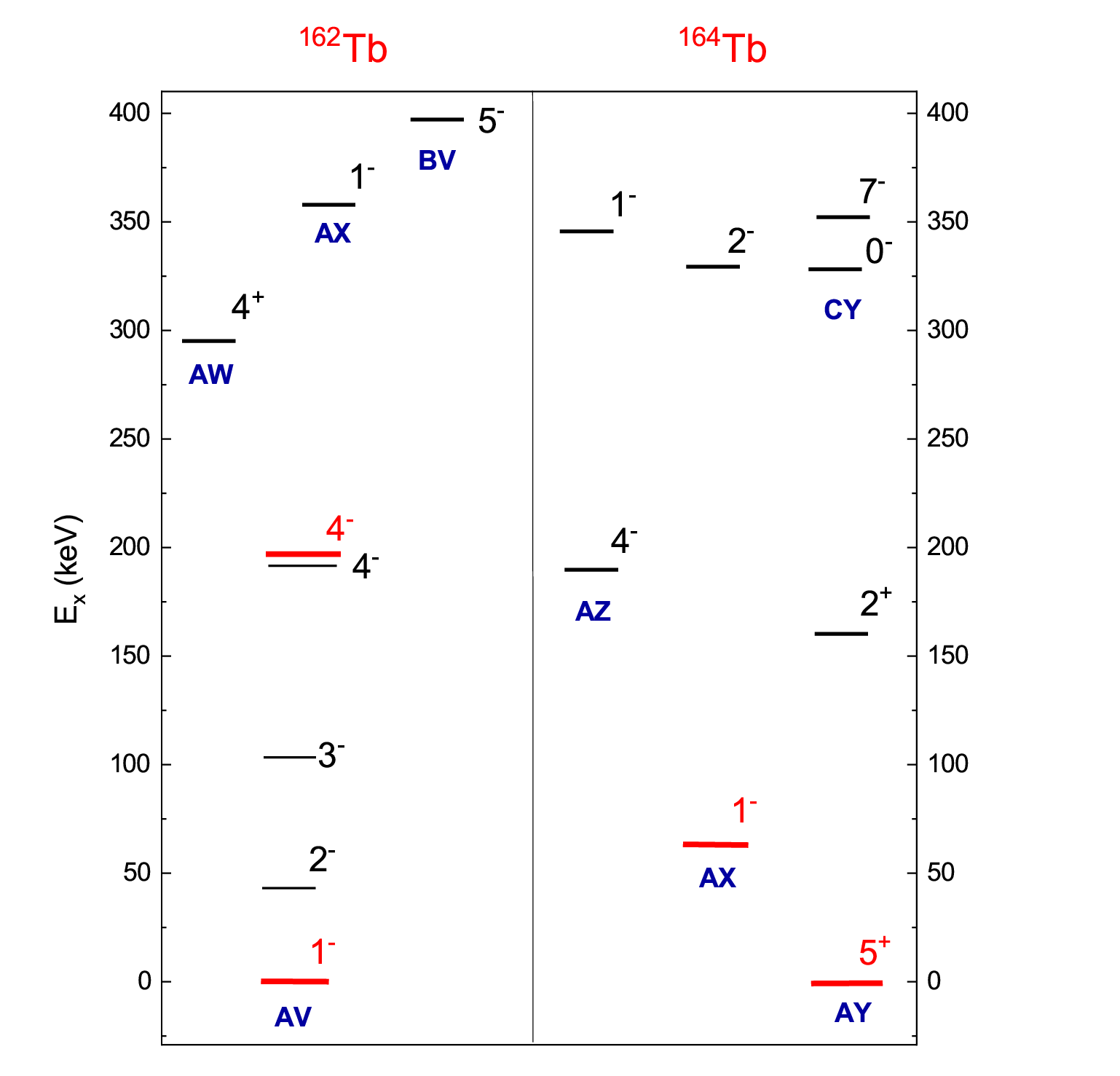}
\caption{Model-calculated energies of the physically admissible 2qp bandheads in $^{162}$Tb (left) and $^{164}$Tb (right) with E$_x<$ 400 keV. Expected low-lying isomers are highlighted in red.} \label{fig2}
\end{figure}

Here, E$_p$ and E$_n$ are experimental single particle excitation energies for proton and neutron orbitals, respectively, taken from the neighboring A$\pm$1 nuclei. 
E$_{rot}$ is the rotational energy correction term and V$_{pn}$ is the p-n residual interaction defined in terms of the GM splitting energy between the triplet and singlet state of the formed doublet (E$_{GM}$) and the Newby odd-even shift (E$_N$) in case of K=0 bandheads. Detailed description of the TQRM formulation can be found in \cite{Jain90,Jain98}. In Fig. \ref{fig2}, we have plotted the model-calculated 2qp bandhead energies for $^{162}$Tb and $^{164}$Tb. 

In $^{162}$Tb, the ground state (gs) GM doublet is formed by (AV): 1$^-$\{ $\pi$3/2$^+$[411$\uparrow$]$\otimes$$\nu$5/2$^-$[523$\downarrow$]\}4$^-$. Our calculations place the spins-antiparallel (J$^\pi$=4$^-$) singlet state of this GM doublet at an energy of E$_x$ $\approx$ 194 keV; all other 2qp bandheads lie above 250 keV as seen in Fig. \ref{fig2}. With $\Delta$J = $\Delta$K = 3, $\gamma$ transitions from this J$^\pi$=4$^-$ level to either the gs J$^\pi$=1$^-$  or its rotational levels will be hindered, rendering an isomeric character to this level. It is noteworthy that our calculated energies of the gs rotational levels are in close agreement with the adopted values \cite{ENSDF}. 
As commented by the ENSDF evaluators, the AME20 \cite{AME20} adopted Q($\beta^-$) value of 2301(22) keV for $^{162}$Tb does not allow the population of the high energy $^{162}$Dy daughter level at E$_x$=2371 keV. 
Kawade \textit{et al.} \cite{KAWADE77}, from their $\beta$-$\gamma$ coincidence studies, reported Q($\beta^-$)= 2580(22) keV.
Going by our estimated level structure, the presence of an independently $\beta$-decaying J$^\pi$=4$^-$ isomer with E$_x$$\thickapprox$200 keV allows the $\beta$-feeding of the 2371 keV daughter level and agrees with the experimentally measured Q($\beta^-$) value for $^{162}$Tb \cite{KAWADE77}. 
The spins of the daughter level, currently listed as J=1$^-$,2,3 from the $\beta$-feeding and $\gamma$ decays to J=2,3 levels \cite{ENSDF}, also support the possibility of $\beta$-feeding from a J$^\pi$=4$^-$ parent state.         

Similarly, in $^{164}$Tb, the (AY): J$^\pi$= 5$^+$ 
\{$\pi$3/2$^+$[411$\uparrow$]$\otimes$$\nu$7/2$^+$[633$\uparrow$]\} is confirmed to be the gs with its singlet partner J$^\pi$=2$^+$ expected around 160 keV.
Orford \textit{et al.} \cite{Orford2020} had proposed this singlet state as the isomer with E$_x$=170 keV.
However, our calculations place the (AX): J$^\pi$= 1$^-$\{$\pi$3/2$^+$[411$\uparrow$]$\otimes$$\nu$1/2$^-$[521$\downarrow$]\} state at E$_x$$\thickapprox$66 keV which could be the possible spin-isomer since the only level to which it can decay is the J$^\pi$= 5$^+$ gs, with a spin difference of $\Delta$J=4. 
Thus our deduced level structure affirms the existence of an isomeric state, but the expected isomer is not the gs singlet partner J$^\pi$=2$^+$ as proposed by Orford \textit{et al.} \cite{Orford2020}. 

In conclusion, our TQRM studies of low-lying level structures in $^{162,164}$Tb reveal the presence of 2qp states that may correspond to the spin-trap isomers predicted by Orford \textit{et al.} \cite{Orford2020}, viz.,

\textbf{$^{162}$Tb$^m$}: 4$^-$\{$\pi$3/2$^+$[411$\uparrow$]$\otimes$$\nu$5/2$^-$[523$\downarrow$]\};E$_x$$\thickapprox$194 keV

\textbf{$^{164}$Tb$^m$}: 1$^-$\{$\pi$3/2$^+$[411$\uparrow$]$\otimes$$\nu$1/2$^-$[521$\downarrow$]\};E$_x$$\thickapprox$66 keV.

In $^{162}$Tb, our model-calculated bandhead energies are in close agreement with existing experimental data for 2qp bandheads upto E$_x$=400 keV. In $^{164}$Tb, where currently no experimental data is available, our TQRM analysis admits eight 2qp bandheads below E$_x$ = 400 keV. The presence of isomers and other low-lying states in both these nuclei need further experimental validation. Dedicated $\beta$ and $\gamma$ spectroscopy studies of $^{162,164}$Tb supported by internal conversion measurements can be carried out, for which our assignments can serve as location guides.

\nocite{*}

\bibliography{apssamp}% Produces the bibliography via BibTeX.

\end{document}